\documentclass[prb,amsmath,amssymb,superscriptaddress,floatfix,twocolumn,showpacs]{revtex4}

\usepackage{bm}
\usepackage{color}
\usepackage{graphicx}
\usepackage{amsmath}
\usepackage{amsfonts}
\usepackage{epsfig}
\usepackage{epsf}
\usepackage{array}

\newcommand{\vect}[1]{\mbox{\boldmath $\rm #1$}}

\begin{document}

\title{Anomalous mass enhancement in strongly-correlated quantum wells}

\author{Satoshi Okamoto}
\altaffiliation{okapon@ornl.gov}
\affiliation{Materials Science and Technology Division, Oak Ridge National Laboratory, Oak Ridge, Tennessee 37831, USA}

\begin{abstract}
Using dynamical-mean-field theory, 
we investigate the electronic properties of quantum wells consisting of a $t^1_{2g}$-electron system with strong correlations. 
The special focus is on the subband structure of such quantum wells. 
The effective mass is found to increase with increase in the value of the bottom of the subband, i.e., 
decrease in the subband occupation number. 
This is due to the combination of Coulomb repulsion, whose effect is enhanced on surface layers, 
and longer-range hoppings. 
We discuss the implication of these results for the recent angle-resolved photoemission experiment on SrVO$_3$ thin films. 
\end{abstract}

\maketitle


Two-dimensional electron gases (2DEGs) realized in a variety of oxide interfaces have been attracting significant interests.%
\cite{Ohtomo02,Ohtomo04}
In particular, 
electronics utilizing oxides with strong correlations would benefit from their rich phase behaviors.\cite{ScienceReview10}
For example, control of the band structure of 2DEGs in transition-metal oxides has been proposed as a way 
to create non-cuprate high-$T_c$ superconductivity.\cite{Chaloupka08} 
Yet, the realization of metallic behavior in few-unit-cell thick oxides remains challenging.\cite{Yoshimatsu10,Liu11} 

Two-dimensional metallic behavior in confined geometry, i.e., in quantum wells (QWs), has been studied for conventional metals. 
Reconstructed band structures or subband dispersion relations in QWs of Ag thin films 
have been confirmed using photoemission spectroscopy.\cite{Evans93,Matsuda02} 
The subband dispersion of 2DEGs realized on the surface of a band insulator SrTiO$_3$ 
was also observed using angle-resolved photoemission spectroscopy (ARPES).\cite{Syro11,Meevasana11}
More recently, Yoshimatsu and coworkers have performed ARPES measurements on QWs 
in thin films of the correlated metal SrVO$_3$.\cite{Yoshimatsu11} 
The subband structures realized in such QWs can be explained reasonably well using a simple tight-binding-type description. 
However, the effective mass of such subbands was found to increase with decreasing binding energy of the subband. 
Since this trend is opposite to what we expect based on the bulk behavior, 
i.e., the effective mass is reduced with decreasing binding energy and decreasing band occupancy, 
the origin of such an anomalous mass enhancement remains to be understood.

In this Rapid Communications, 
we analyze model QWs consisting of a $t_{2g}^1$ electron system as experimentally considered by Yoshimatsu {\it et al}.
We employ layer dynamical-mean-field theory (DMFT) with the exact-diagonalization impurity solver.\cite{Georges96,Caffarel94} 
In correlated QWs, a smaller coordination number on surface layers induces larger mass enhancement than in 
the bulk region.\cite{Potthoff99a,Potthoff99b,Schwieger03,Liebsch03,Ishida06} 
This brings about the anomalous subband-dependent mass enhancement; 
the effective mass becomes larger with decreasing subband binding energy or depopulation of the subband. 
With the additional effect coming from the longer-range hopping, 
the subband-dependent mass enhancement was found to increase dramatically. 
We argue that the anomalous mass enhancement reported for thin films of SrVO$_3$ is due to 
strong correlations and long-range transfer integral.

We consider the three-band Hubbard model involving $t_{2g}$ electrons, 
$H = H_{\rm band} + \sum_{\vect r} H_{\rm loc}(\vect r)$. 
The first term describes the non-interacting part of the system as 
\begin{eqnarray}
H_{\rm band} = - \sum_{\tau, \sigma} \sum_{\vect r, \vect r'} t^\tau_{\vect r \vect r'} 
d^\dag_{\vect r \tau \sigma} d_{\vect r' \tau \sigma} ,
\end{eqnarray}
where $d_{\vect r \tau \sigma}$ stands for the annihilation operator 
for an electron at site $\vect r$ in orbital $\tau$ with spin $\sigma$, 
and $t^\tau_{\vect r \vect r'}$ is the hopping integral between orbital $\tau$ at sites $\vect r$ and $\vect r'$. 
For the orbital $\tau = xy$, we take the nearest neighbor hoppings 
$t^\tau_{\vect r \vect r'} = t_\pi$ for $\vect r'=\vect r \pm \hat {\vect x} (\hat {\vect y})$ and 
$t^\tau_{\vect r \vect r'} = t_\delta$ for $\vect r'=\vect r \pm \hat {\vect z}$, 
and the second-neighbor hopping $t^\tau_{\vect r \vect r'} = t_{\sigma'}$ for 
$\vect r' = \vect r \pm \hat {\vect x} \pm \hat {\vect y}$. 
Here, $\hat {\vect x} (\hat {\vect y},\hat {\vect z})$ is the unit vector along the $x (y,z)$ direction. 
The hopping parameters for the other orbitals are given by interchanging the coordinates $x$, $y$, and $z$ accordingly. 
Parameter values are taken from density functional theory results as
$t_\pi = 0.281$, $t_\delta = 0.033$, and $t_{\sigma'} = 0.096$ (all in eV).\cite{Pavarini05} 
$H_{\rm loc}$ describes the local interaction as 
\begin{equation}
H_{\rm loc} = 
\frac{1}{2} \!\!\! 
\sum_{{\scriptstyle \tau, \tau', \tau''}\atop{\scriptstyle \tau''',\sigma, \sigma'}} 
\!\!\! 
U^{\tau \tau' \tau'' \tau'''} 
d_{\tau \sigma}^\dag d_{\tau' \sigma'}^\dag d_{\tau''' \sigma'} d_{\tau'' \sigma}
- \mu \sum_{\tau, \sigma} d^\dag_{\tau \sigma} d_{\tau \sigma}. \label{eq:Hloc} 
\end{equation}
Here, the site index $\vect r$ is suppressed for simplicity, and $\mu$ is the chemical potential. 
Since we consider $t_{2g}$ electron systems, we assume the relation 
$U=U'+2J$, where 
$U=U^{\tau \tau \tau \tau}$ (intraorbital Coulomb), 
$U'=U^{\tau \tau' \tau \tau'}$ (interorbital Coulomb), 
$J=U^{\tau \tau' \tau' \tau}$ (interorbital exchange) 
$= U^{\tau \tau \tau' \tau'}$ (interorbital pair transfer) for $\tau \ne \tau'$, 
and other components are absent.\cite{Sugano70} 
As in Ref.~\onlinecite{Yoshimatsu11}, 
we consider QWs in which a finite number of correlated layers stack along the $z$ direction 
with the open-boundary condition 
and the periodic-boundary condition along the $x$ and $y$ directions. 

\begin{figure}[tbp]
\includegraphics[width=0.8\columnwidth,clip]{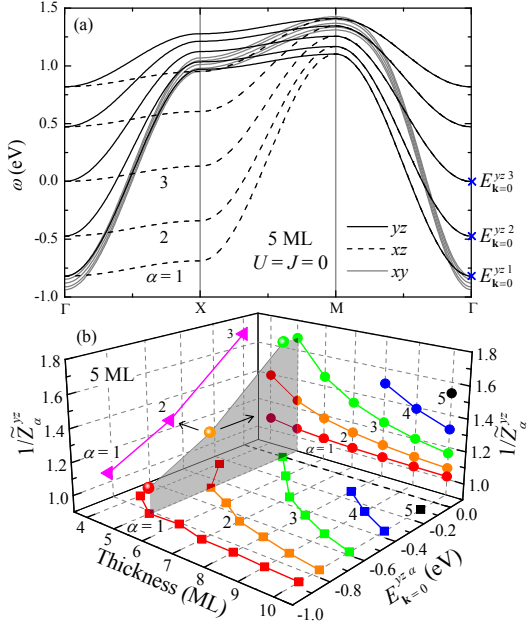}
\caption{(Color online) (a) Dispersion relation as a function of momentum for a non-interacting 5-ML-thick quantum well. 
Crosses indicate $E^{yz \, \alpha}_{\vect k = 0}$ (binding energy at $\vect k=0$ times $-1$). 
(b) Three-dimensional plot of subband quasiparticle weight $\widetilde Z_\alpha^{yz}$, $E^{yz \, \alpha}_{\vect k = 0}$, and QW thickness. 
For each QW, $\widetilde Z_\alpha^{yz}$ is defined at $E^{yz \, \alpha}_{\vect k = 0} < 0$. 
When projected on the left (right) vertical plane, $\widetilde Z_\alpha^{yz}$ is given as a function of 
$E^{yz \, \alpha}_{\vect k = 0}$ (QW thickness). 
}
\label{fig:noninterac}
\end{figure}

Before going into the detailed analysis taking into account the correlation effects, 
let us first discuss the low-energy electronic behavior focusing on quasiparticle bands. 
For this purpose, we consider the following effective one-dimensional Schr{\"o}dinger equation\cite{Okamoto04b} 
\begin{eqnarray}
\Bigl[Z^\tau_z \bigl\{\tilde \varepsilon^\tau_{\vect k} - \mu + {\rm Re} \Sigma^\tau_z (0) \bigr\}\delta_{z,z'} 
+ \sqrt{Z^\tau_z Z^\tau_{z'}} t^\tau_{\vect k} \delta_{z,z' \pm 1} \Bigr] \varphi^{\tau \alpha}_{z' \vect k} 
\nonumber \\
= E^{\tau \alpha}_{\vect k} \varphi^{\tau \alpha}_{z \vect k}, \quad
\label{eq:1d_schroedinger}
\end{eqnarray}
where $\Sigma^\tau_z(\omega)$ is the electron self-energy at orbital $\tau$ in layer $z$ computed in layer DMFT, 
and $Z^\tau_z$ is the layer-dependent quasiparticle weight defined by 
$Z^\tau_z = \{1 - {\rm Re} \partial_\omega \Sigma^\tau_z(\omega)|_{\omega = 0}\}^{-1}$. 
$\tilde \varepsilon^\tau_{\vect k}$ is the in-plane dispersion and $t^\tau_{\vect k}$ is the out-of-plane hopping element for orbital $\tau$ 
with in-plane momentum $\vect k=(k_x,k_y)$. 
For $\tau = yz$, these are explicitly given by 
$\tilde \varepsilon^{yz}_{\vect k} = - 2 t_\pi \cos k_y - 2 t_{\sigma'} \cos k_x$ and 
$t^{yz}_{\vect k} = - t_\pi - 2 t_\delta \cos k_y$. 
$\alpha$ labels the subband with the eigenfunction $\varphi^{\tau \alpha}_{z \vect k}$
in increasing order of the subband energy $E^{\tau \alpha}_{\vect k}$. 
As an example, the energy eigenvalue $E_{\vect k}^{\tau \alpha}$ for 5-ML-thick non-interacting ($Z^\tau_z=1$) QW 
is plotted in Fig.~\ref{fig:noninterac}~(a) (ML indicates monolayer). 
We notice that subbands originating from $yz (xz)$ orbitals are not parallel, while $xy$ subbands are. 
This is because the second-neighbor hopping between neighboring layers induces $\vect k$ dependence in 
the out-of-plane hopping $t^{yz (xz)}_{\vect k}$. 
For orbital $yz (xz)$, $t^\tau_{\vect k} =- t_\pi - 2 t_\delta \cos k_{y (x)}$ and, therefore, 
the subband separation becomes large when $k_{y (x)}$ approaches 0. 
As a result, the Fermi velocity 
of high-energy (less-populated) bands 
becomes small as if 
{\it the effective mass is enhanced}. 

The low-energy electronic behaviors of correlated QWs are governed by the quasiparticle subbands. 
The correlation effects enter as the quasiparticle weight 
of the subband. 
Using the solution of Eq.~(\ref{eq:1d_schroedinger}), the subband-dependent quasiparticle weight is given by\cite{Okamoto04b,Ruegg07} 
\begin{eqnarray}
Z^\tau_\alpha = \sum_z Z^\tau_z \Bigl| \varphi^{\tau \alpha}_{z \, \vect k = \vect k^\alpha_F} \Bigr|^2. 
\label{eq:Z}
\end{eqnarray}
From Eq.~(\ref{eq:Z}), it is clear that the subband quasiparticle weight becomes unity in the absence of correlations, 
i.e., $Z^\tau_z=1$ leads to 
$\sum_z \bigl| \varphi^{\tau \alpha}_{z \, \vect k = \vect k^\alpha_F} \bigr|^2=1$ (normalization of the quasiparticle eigenfunction). 
Another important quantity is the {\it effective quasiparticle weight} defined by 
\begin{eqnarray}
\widetilde Z^\tau_\alpha = \left. 
\frac{\partial_{\vect k} E^{\tau \alpha}_{\vect k}}{\partial_{\vect k} \varepsilon^\tau_{\vect k 0}} 
\right|_{\vect k = \vect k^{\tau \alpha}_F}. 
\end{eqnarray}
Here, $\vect k^{\tau \alpha}_F$ is the Fermi momentum for the $\alpha$th subband, 
and $\varepsilon^\tau_{\vect k k_z}$ is the bulk dispersion. 
For $\tau = yz$, we have 
$\varepsilon^{yz}_{\vect k k_z} = \tilde \varepsilon^{yz}_{\vect k} + 2 t^{yz}_{\vect k} - 2 t_\pi \cos k_z$.
Thus, $\widetilde Z^\tau_\alpha$ measures the change in the Fermi velocity with respect to its bulk value. 
In Ref.~\onlinecite{Yoshimatsu11}, $\widetilde Z^\tau_\alpha$ was used to discuss the mass enhancement.

Because of the $\vect k$ dependence of the interlayer hopping matrix $t^\tau_{\vect k}$, 
$\widetilde Z$ can be smaller than unity even without correlations. 
Figure~\ref{fig:noninterac}~(b) summarizes the results for $E_{\vect k=0}^{yz \, \alpha}$ and $1/\widetilde Z$ 
for non-interacting QWs with thickness varied from 4 to 10. 
As the QW becomes thin, $E_{\vect k=0}^{yz \, \alpha}$ increases and the number of occupied subbands is reduced (see the basal plane). 
For a 5-ML-thick QW, $1/\widetilde Z_\alpha^{yz}$ is projected in the left vertical plane 
thus is shown as a function of $E_{\vect k=0}^{yz \, \alpha}$ 
(binding energy times $-1$). 
As $E_{\vect k=0}^{yz \, \alpha}$ approaches 0, $1/\widetilde Z_\alpha^{yz}$ is increased. 
This trend can be seen in all QWs studied 
(see the projection of $1/\widetilde Z_\alpha^{yz}$ on the right vertical plane). 
A similar trend was reported experimentally. 
However, $1/\widetilde Z^\tau_\alpha$ is enhanced from $\sim 1$ to $\sim 1.7$, so is at most 70~\%. 
Therefore, the band effect alone does not account for the large mass enhancement reported in Ref.~\onlinecite{Yoshimatsu11}, where 
$1/\widetilde Z^\tau_\alpha$ varies from $\sim 1.7$ to $\sim 4.5$. 
The experimental enhancement in $1/\widetilde Z^\tau_\alpha$ is nearly 300~\%, 
and $1/\widetilde Z^\tau_\alpha$ at the largest binding energy is already $\sim 70$~\% larger than the band mass. 
These results indicate the influence of the correlation effects.

In order to see the effect of correlations rather quantitatively, 
here we employ layer DMFT, whose self-consistency condition is closed by\cite{Potthoff99a,Potthoff99b,Schwieger03,Okamoto04b}
\begin{equation}
G^\tau_{z} (\omega) = \!\! \int \!\! \frac{d^2 k}{(2 \pi)^2} G^\tau_{zz} (\vect k,\omega).
\end{equation}
Here, $G^\tau_{z}$ is the local Green's function on layer $z$, 
and the lattice Green's function matrix on the right-hand side is given 
as a function of $\vect k$ and the $z$-axis coordinate as 
$\hat G (\vect k, \omega) = \bigl[ (\omega + \mu) \vect 1 - \hat H_{\rm band} (\vect k) - \hat \Sigma(\vect k, \omega) \bigr]^{-1}$. 
The hopping matrix $\hat H_{\rm band} (\vect k)$ is given by 
an in-plane Fourier transformation of $H_{\rm band}$  as 
$\hat H_{\rm band} (\vect k) = (\tilde \varepsilon^\tau_{\vect k} \delta_{z,z'} + t^\tau_{\vect k} \delta_{z,z' \pm 1})\delta_{\tau,\tau'}$. 
The self-energy matrix is approximated as 
$\hat \Sigma(\vect k, \omega) = \Sigma^\tau_z(\omega) \delta_{z,z'} \delta_{\tau,\tau'}$. 
The local self-energy is obtained by solving the effective impurity model defined 
by the local interaction term coupled with an effective medium. 
In this study, we use the exact diagonalization impurity solver with the Arnoldi algorithm.\cite{Lehoucq97,Perroni07} 
Here, the effective medium is approximated as a finite number of bath sites, and 
the impurity Hamiltonian is given by
\begin{equation}
H_{\rm imp} = H_{\rm loc} + \sum_{i, \tau, \sigma} \varepsilon_{i \tau} c^\dag_{i \tau \sigma} c_{i \tau \sigma} 
+ \sum_{i, \tau, \sigma} \Bigl(v_{i \tau} c^\dag_{i \tau \sigma} d_{\tau \sigma} + h.c. \Bigr). 
\end{equation}
$c_{i \tau \sigma}$ is the annihilation operator of an electron at the $i$th bath site with 
potential $\varepsilon_{i \tau}$ and hybridization strength with the impurity orbital $\tau$ denoted by $v_{i \tau}$. 
Because of the exponentially growing Hilbert space with respect to the numbers of orbitals and electrons, 
we consider two bath sites per correlated orbital, i.e., $i=1,2$. 
In our numerical simulations, 
we use temperature $T=10^{-2}$~eV to retain low-energy states with Boltzmann factors larger than $10^{-6}$ 
and consider only paramagnetic solutions.

\begin{figure}[tbp]
\includegraphics[width=0.8\columnwidth,clip]{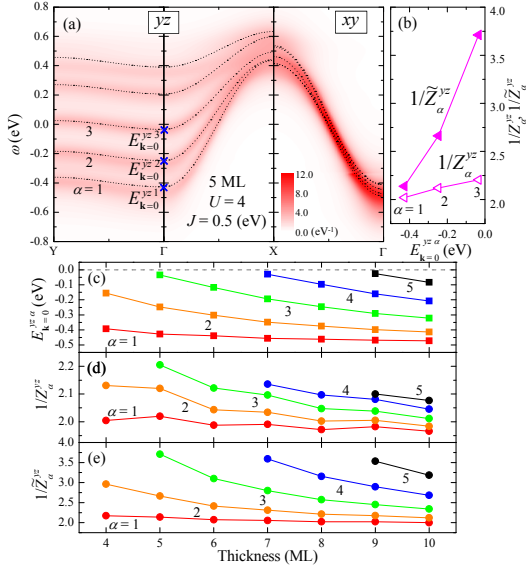}
\caption{(Color online) (a) Orbitally-resolved spectral function as a function of momentum and frequency 
for the interacting 5-ML-thick QW with $U=4$ and $J=0.5$~eV. 
Dotted lines indicate the solution of Eq.~(\ref{eq:1d_schroedinger}), $E^{\tau \alpha}_{\vect k}$, 
obtained using the layer DMFT result for the quasiparticle weight $Z^\tau_z$. 
Note that $yz$ and $xz$ bands are symmetric with respect to $X \leftrightarrow Y$. 
(b) Subband quasiparticle weights $Z_\alpha^{yz}$ and $\widetilde Z_\alpha^{yz}$ as functions of $E^{yz \, \alpha}_{\vect k = 0}$. 
(c) $E^{yz \, \alpha}_{\vect k}$, 
(d) quasiparticle weight $Z_\alpha^{yz}$, and 
(e) effective quasiparticle weight $\widetilde Z_\alpha^{yz}$ as functions of the thickness of QWs. 
Plots (b)--(e) are generated using the method displayed in Fig.~\ref{fig:noninterac} (b). 
See also Ref.~\onlinecite{supplement} (Fig.~S1). 
}
\label{fig:interac}
\end{figure}

Figure \ref{fig:interac}~(a) shows the results for the orbitally-resolved spectral function 
$A^\tau (\vect k, \omega)=-\frac{1}{\pi} \sum_z {\rm Im} G^\tau_{zz} (\vect k, \omega)$
as well as $E^{\tau \alpha}_{\vect k}$ as dotted lines 
for a 5-ML-thick interacting QW with $U=4$ and $J=0.5$~(both in eV). 
For $A^\tau (\vect k, \omega)$, the self-energy is extrapolated to the real axis using the Pad{\'e} approximation.\cite{Perroni07} 
In comparison with the non-interacting case, 
the bandwidth is reduced by about 50~\% due to correlations. 
We notice that $E^{\tau \alpha}_{\vect k}$ reproduces the peak positions of $A^\tau (\vect k, \omega)$ fairly well. 
There are five subbands for both $yz$ and $xy$, 
but those in the latter are indistinguishable because all subbands are located within the range of $2t_\delta \sim 0.07$~eV. 
Thus, we focus on $yz$ subbands in the following discussion. 
Using the same procedure as in Fig.~\ref{fig:noninterac}~(b), 
we analyze $E^{yz \alpha}_{\vect k=0}$ and the mass enhancements $1/Z^{yz}_\alpha$ and $1/\widetilde Z^{yz}_\alpha$. 
Figure~\ref{fig:interac}~(b)shows plots of $1/Z^{yz}_\alpha$ and $1/\widetilde Z^{yz}_\alpha$ as functions of $E^{yz \alpha}_{\vect k=0}$ 
for a 5-ML-thick QW. 
At the largest binding energy, both $1/Z^{yz}_\alpha$ and $1/\widetilde Z^{yz}_\alpha$ are about 2, 
the mass enhancement expected in the bulk region. 
Figures~\ref{fig:interac}~(c), \ref{fig:interac}~(d), and \ref{fig:interac}~(e) summarize 
$E_{\vect k = 0}^{yz \, \alpha}$, $1/Z_\alpha^{yz}$, and $1/\widetilde Z_\alpha^{yz}$, respectively, for 
interacting QWs with thickness varied from 4 to 10.  
Although $1/Z^{yz}_\alpha$ shows an increase with increasing $E^{yz \, \alpha}_{\vect k=0}$, 
it is only from $\sim 2$ to $\sim 2.2$. 
On the other hand, $1/\widetilde Z^{yz}_\alpha$ shows a rather steep increase from $\sim 2$ to $\sim 3.8$, 
as reported experimentally.\cite{Yoshimatsu11}

\begin{figure}[tbp]
\includegraphics[width=0.8\columnwidth,clip]{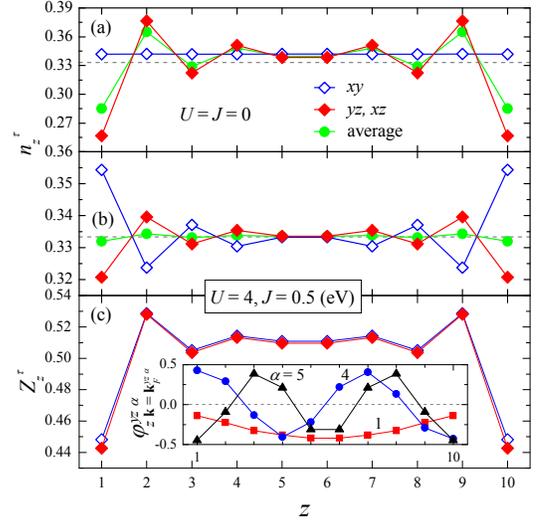}
\caption{(Color online) (a) Orbitally-resolved occupation number as a function of position $z$ for a non-interacting 10-ML-thick QW. 
(b) Same as (a) for the interacting model and (c) the local quasiparticle weight as a function of $z$. 
Layer DMFT with $U=4$ and $J=0.5$~eV is used for (b) and (c). 
Inset: Quasiparticle wave functions $\varphi^{yz \, \alpha}_{z \vect k = \vect k_F^{yz \, \alpha}}$. 
}
\label{fig:densityZ}
\end{figure}

Aside from the quantitative difference, non-interacting QWs and interacting QWs behave quite similarly. 
As a small but clear difference, some of interacting QWs have a larger number of occupied $yz$ subbands, 7 and 9-ML-thick QWs. 
This is caused by the different orbital polarization. 
As shown in Figs.~\ref{fig:densityZ}~(a) and \ref{fig:densityZ}~(b), 
non-interacting QWs have larger orbital polarization with smaller occupancy in 
the $yz$ and $xz$ orbitals on surface layers than interacting QWs. 
This is because these bands have a quasi-one-dimensional character on the surface layers, with a reduced effective bandwidth. 
The average charge density also shows Friedel-type oscillatory behavior with respect to $z$. 
On the other hand, in the interacting case, the orbital polarization and the charge redistribution are significantly suppressed 
because the charge susceptibility is suppressed near integer fillings. 
This behavior was found to be insensitive to the choice of the interaction strength as shown in Ref.~\onlinecite{supplement} (Fig.~S2). 
Therefore, the effect of charge relaxation is expected to be small, in contrast to LaTiO$_3$/SrTiO$_3$ heterostructures.

Figure~\ref{fig:densityZ}~(c) shows the position-dependent quasiparticle weight $Z_z^\tau$, 
and its inset the quasiparticle eigenfunctions for a $yz$ electron at the Fermi level. 
Strong mass renormalization takes place in surface layers where the coordination number is smaller.%
\cite{Potthoff99a,Potthoff99b,Schwieger03,Liebsch03,Ishida06} 
In the current case, $Z_z^\tau$ is $\sim 0.43 (0.51)$ on surface layers (in the bulk region at $z=5$); 
thus there is about 15~\% stronger mass renormalization on the surface. 
This small difference comes from the fact that SrVO$_3$ is not so strongly correlated. 
With increasing $U$, $Z_z^\tau$ on surfaces are more strongly renormalized [see Ref.~\onlinecite{supplement}, Fig.~S3 (a)]. 
Since an eigenfunction with larger $\alpha$ has larger weight on the surface layers, 
the effective mass of such a subband is more strongly renormalized. 
But, the renormalization of $Z^{yz}_\alpha$, up to $\sim 10$~\%, 
is smaller than that of $Z_z^\tau$ because of the interlayer hybridization. 
%
%
The additional enhancement in $1/\widetilde Z^{yz}_\alpha$ is caused by the momentum dependent inter-layer hoppings, as discussed earlier.

\begin{figure}[tbp]
\includegraphics[width=0.8\columnwidth,clip]{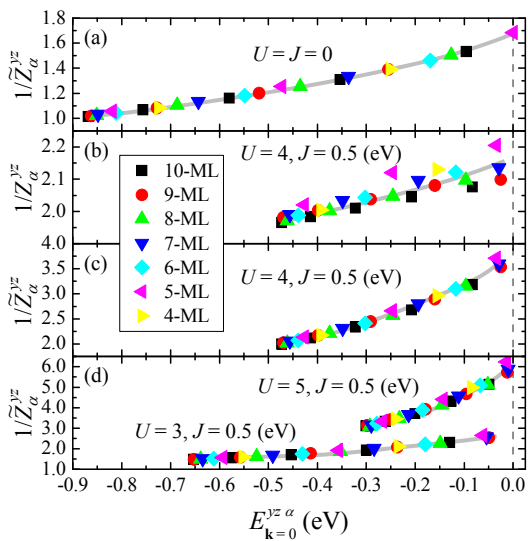}
\caption{(Color online) Subband mass enhancement. 
(a) Effective mass enhancement $1/\widetilde Z^{yz}_\alpha$ for the non-interacting model as a function of
$E_{\vect k=0}^{yz \, \alpha}$. 
(b) Mass enhancement $1/Z^{yz}_\alpha$ and (c) $1/\widetilde Z^{yz}_\alpha$ for the interacting model 
with $U=4$ and $J=0.5$~(eV).  
(d) $1/\widetilde Z^{yz}_\alpha$ for the interacting model with the parameter values indicated. 
Gray bold lines are guides to the eye. 
}
\label{fig:massenhancement}
\end{figure}

Figure~\ref{fig:massenhancement} summarizes the mass enhancement as a function of the position of the bottom of the subband. 
The enhancement in $1/\widetilde Z^{yz}_\alpha$ is rather small for non-interacting QWs 
because it comes from the small hopping parameter $t_{\sigma'}$. 
The mass enhancement $1/Z^{yz}_\alpha$ originating purely from the correlation effects also shows 
rather small dependence on $E_{\vect k=0}^{yz \, \alpha}$. 
On the other hand, $1/\widetilde Z^{yz}_\alpha$ shows strong $E_{\vect k=0}^{yz \, \alpha}$ dependence 
because both the band and correlation effects are included. 
The effective mass enhancement $1/\widetilde Z^{yz}_\alpha$ somewhat depends on the correlation strength, 
as shown in Fig.~\ref{fig:massenhancement}~(d). 
Comparison of $1/ Z^{yz}_\alpha$ for different interaction strengths is presented in Ref.~\onlinecite{supplement} [Figs.~S3~(b)--S3~(c)].
With a reasonable parameter set, the experimentally reported mass enhancement can be semiquantitatively reproduced.

We notice that the number of subbands is overestimated by $\sim 1$ for interacting QWs compared 
with the experimental observation.\cite{Yoshimatsu11} 
A possible explanation for this discrepancy is that, in the experiment of Ref.~\onlinecite{Yoshimatsu11}, 
the surface layer is made of VO$_2$,  
so that the symmetry and the valence state of surface V ions greatly deviate from those in the bulk. 
Also, we cannot exclude the possibility of surface lattice relaxation by which conduction electrons 
are strongly localized on the surface layer. 
In these cases, the surface V sites would not contribute to the ARPES spectrum near the Fermi level as do those in the bulk. 
Detailed study including these effects would be necessary to fully understand the nature of SrVO$_3$ QWs 
including the dimensional crossover and the metal-insulator transition.\cite{Yoshimatsu10,Yoshimatsu11} 
Yet, 
the present study provides a reasonable account for 
the anomalous mass enhancement reported for SrVO$_3$ thin films.

Summarizing, using dynamical-mean-field theory, 
we investigated the electronic properties of correlated quantum wells consisting of a $t_{2g}^1$-electron system. 
The special focus is on the subband structure of such quantum wells. 
The subband effective mass was found to increase with decreasing band occupancy as reported for SrVO$_3$ thin films. 
The present theory provides a reasonable account for this observation 
as the combined effect of Coulomb repulsion, whose effect is enhanced on surface layers, 
and longer-range hoppings. 
Inclusion of these two effects is essential to correctly interpret experimental observations. 

The author thanks H. Kumigashira, K. Yoshimatsu, and A. Fujimori for valuable discussions and for sharing experimental data prior to publication. 
The author is grateful to V. R. Cooper for discussion and to C. G. Baker for his advice on coding. 
This work was supported by the U.S. Department of Energy, 
Office of Basic Energy Sciences, Materials Sciences and Engineering Division.

%
%
%
%
%
\vspace*{0.4cm}
%
%
{\large \bf Supplementary material: Anomalous mass enhancement in strongly-correlated quantum wells}\\
%
%

\renewcommand{\thetable}{S\Roman{table}}
\renewcommand{\thefigure}{S\arabic{figure}}
\renewcommand{\thesubsection}{S\arabic{subsection}}
\renewcommand{\theequation}{S\arabic{subsection}.\arabic{equation}}

\setcounter{secnumdepth}{3}

\setcounter{equation}{0}
\setcounter{figure}{0}

For interacting quantum wells (QWs), each subband has the quasiparticle weight $Z_\alpha^\tau$ defined in Eq.~(4) 
and the effective quasiparticle weight $\widetilde Z_\alpha^\tau$ defined in Eq.~(5). 
Figure S1 shows three-dimensional plots of (a) $\widetilde Z_\alpha^{yz}$ and (b) $\widetilde Z_\alpha^{yz}$ 
as functions of $E_{\vect k=0}^{yz \alpha}$ for 4--10-ML thick QWs with $U=4$ and $J=0.5$~eV. 
On the basal planes, $E_{\vect k=0}^{yz \alpha}$ is plotted. 
When $Z_\alpha^\tau (\widetilde Z_\alpha^\tau)$ is projected in the left (right) vertical plane, 
it is given as a function of $E_{\vect k=0}^{yz \alpha}$ (QW thickness). 

\begin{figure}[tbp]
\includegraphics[width=0.8\columnwidth,clip]{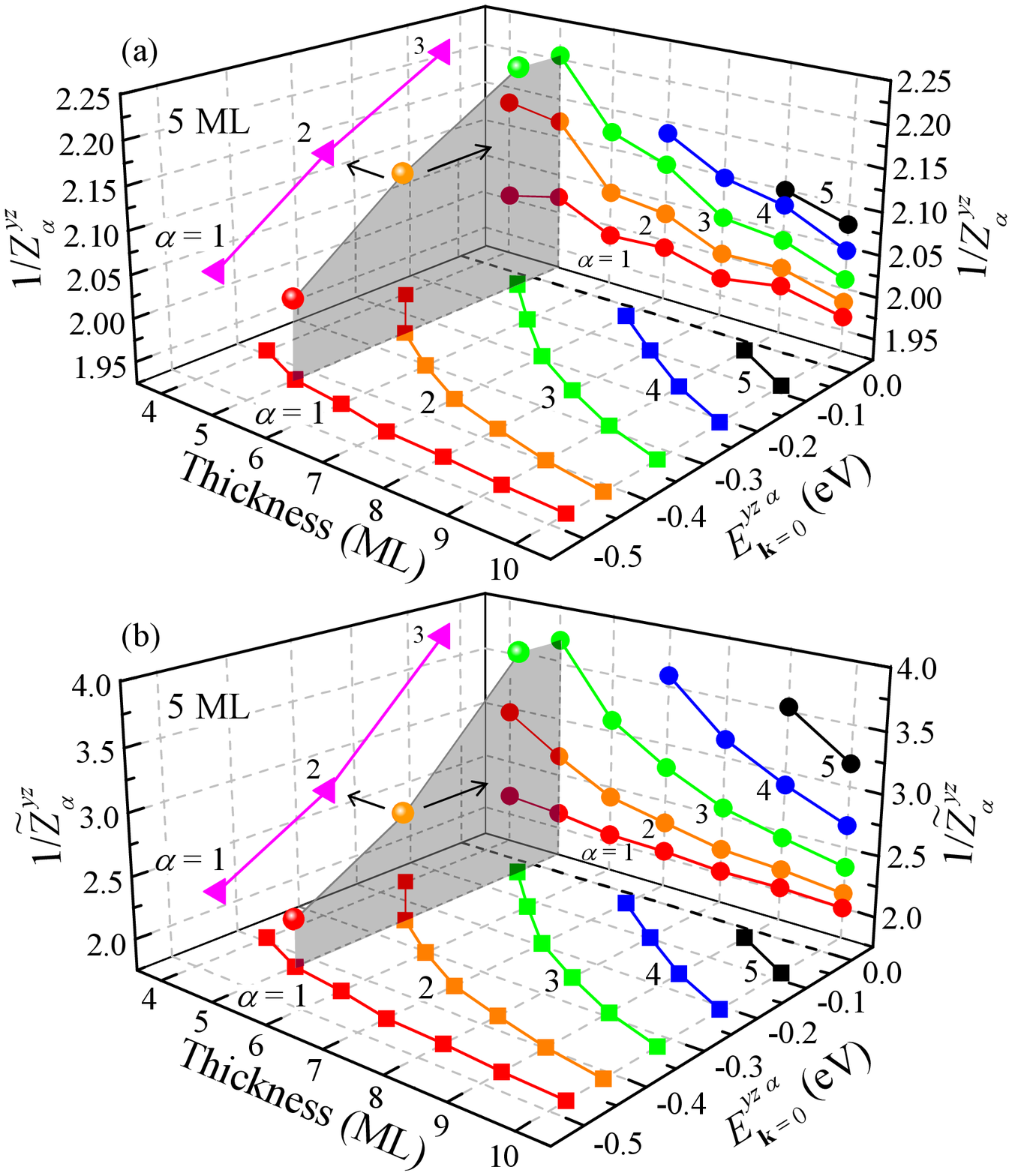}
\caption{Three-dimensional plot of $E_{\vect k=0}^{yz \alpha} < 0$, 
QW thickness, and $Z_\alpha^{yz}$ (a) and $\widetilde Z_\alpha^{yz}$ (b). 
Here, the interaction strength is taken as $U=4$ and $J=0.5$~eV. 
For each QW, $\widetilde Z_\alpha^{yz}$ is defined at $E^{yz \, \alpha}_{\vect k = 0} < 0$. 
When projected on the left (right) vertical plane, 
$\widetilde Z_\alpha^{yz}$ is given as a function of $E^{yz \, \alpha}_{\vect k = 0}$ (QW thickness). 
}
\label{fig:figS1}
\end{figure}

Figure~S2 summarizes the subband effective mass $1/Z_\alpha^{yz}$ as a function of $E_{\vect k=0}^{yz \alpha}$, i.e., the binding energy times $-1$; 
$U=3$ and $J=0.5$ (top panel), $U=4$ and $J=0.5$ [middle panel, same as Fig. 4 (b)], $U=5$ and $J=0.5$~eV (bottom panel).  
With increasing $U$, $1/Z_\alpha^{yz}$ is increased. 
At the same time, $1/Z_\alpha^{yz}$ at 
larger $E_{\vect k=0}^{yz \alpha}$ is enhanced more strongly. 

\begin{figure}[tbp]
\includegraphics[width=0.8\columnwidth,clip]{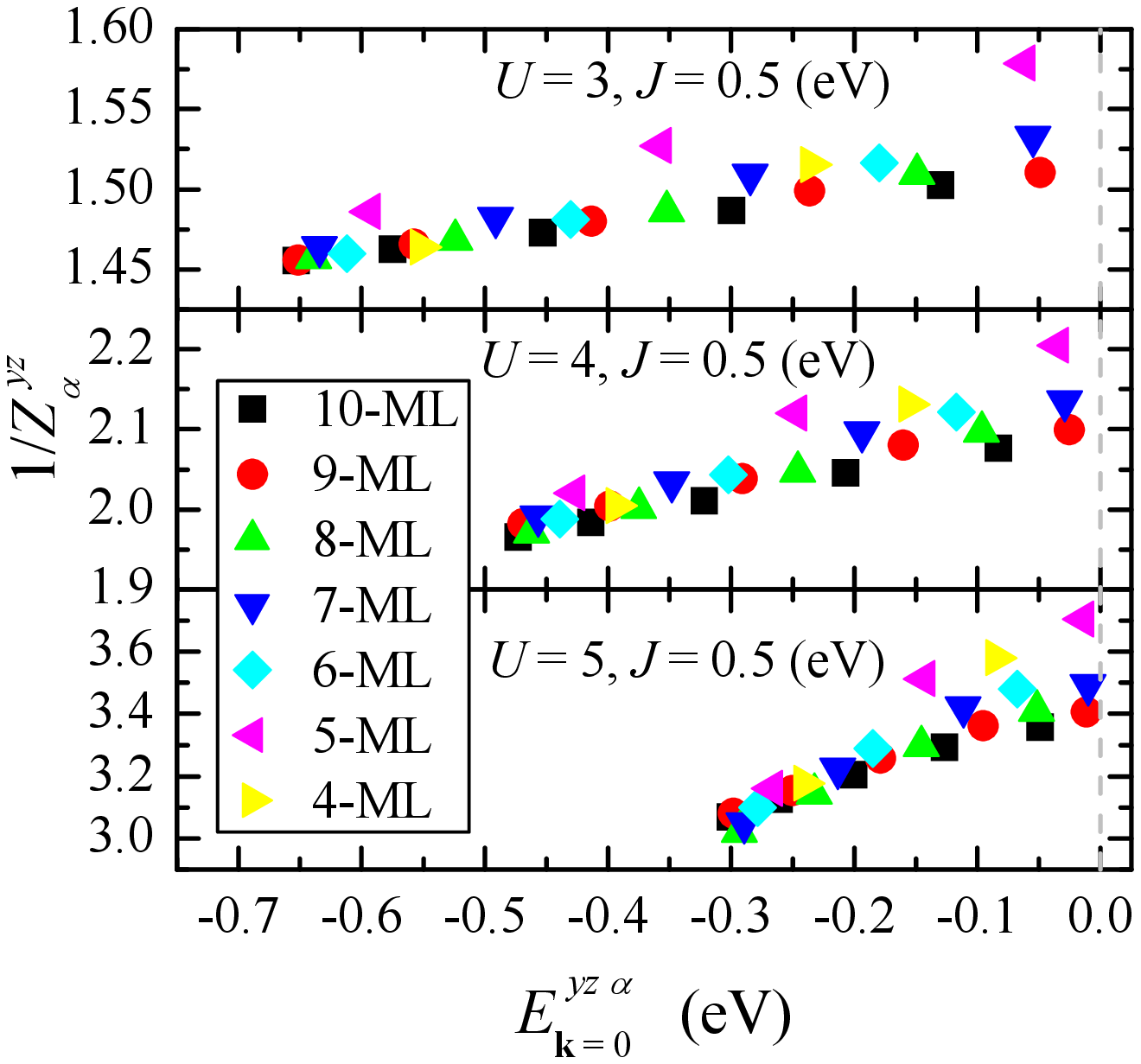}
\caption{Subband quasiparticle weight $Z_\alpha^{yz}$ as a function of $E_{\vect k=0}^{yz \alpha}$ (the binding energy times $-1$) 
with different interaction strengths indicated. 
}
\label{fig:figS2}
\end{figure}

Figure~S3 (a) shows the quasiparticle weight $Z_z^\tau$ as a function of layer index $z$ 
for 10-ML-thick QWs with different interaction strengths. 
With the increase of $U$, $Z_z^\tau$ on surface layers is more strongly renormalized. 

Figures~S3 (b)--(c) show the orbitally resolved charge density as a function of layer index $z$ 
for 10-ML-thick interacting QWs. 
The charge redistribution, deviation of the average density from 1/3, is suppressed with the increase of $U$.

\begin{figure}[tbp]
\includegraphics[width=0.8\columnwidth,clip]{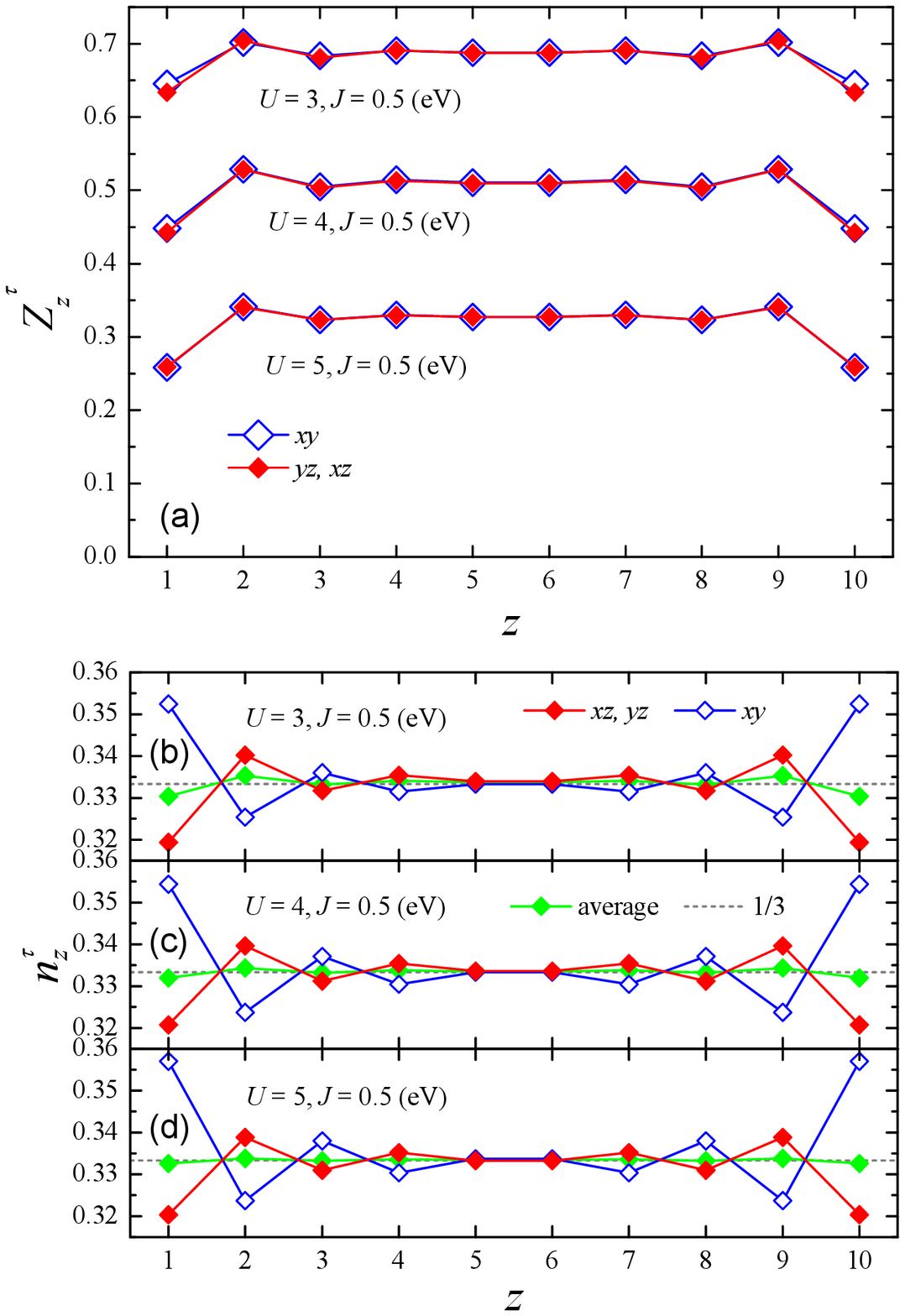}
\caption{(a) Quasiparticle weight $Z_z^\tau$ with $\tau=xy,xz,yz$ 
as a function of layer index $z$ for 10-ML-thick QWs with different interaction strengths indicated. 
With increasing $U$, $Z_z^\tau$ is renormalized, and the renormalization becomes stronger on surface layers at $z=1$ and 10. 
Orbitally resolved charge density as a function of $z$ for 10-ML-thick QWs with 
$U=3$ and $J=0.5$~eV (b), $U=4$ and $J=0.5$~eV (c), and $U=5$ and $J=0.5$~eV (d). 
The average charge density is also shown. Gray dotted lines indicate 1/3. 
}
\label{fig:figS3}
\end{figure}

\end{document}